\documentclass[aps,twocolumn,showpacs,letterpaper,prb]{revtex4}

\usepackage{graphicx}

\begin{document}

\def\asfsix{(TMTTF)$_2$AsF$_6$ }

\def\tco{$T_{CO}$}

\def\tsp{$T_{SP}$}

\def\carbon{$^{13}$C}

\def\as{$^{75}$As}

\def\invtone{$T_1^{-1}$}

\def\iTone{$T_1^{-1}$}

\def\spf{(TMTTF)$_2$PF$_6$}

\def\sasf{(TMTTF)$_2$AsF$_6$}

\def\ssbf{(TMTTF)$_2$SbF$_6$}

\def\Tco{T$_{CO}$}

\def\Tsdw{T$_{SDW}$}

\def\Tsp{T$_{SP}$}

\def\a{\textbf{a}}

\def\b{\textbf{b'}}

\def\c{\textbf{c*}}

\title{Competition and coexistence of bond and charge orders in (TMTTF)$_2$AsF$_6$}
\author{F. Zamborszky$^{1,2}$, W. Yu$^1$, W. Raas$^1$, S. E. Brown$^1$, B. Alavi$^1$, 
C. A. Merlic$^3$, A. Baur$^3$}
\affiliation{$^1$ Department of Physics and Astronomy, UCLA, Los Angeles, California 
90095-1547}
\affiliation{$^2$ Los Alamos National Laboratory, Los Alamos, New Mexico 87545}
\affiliation{$^3$ Department of Chemistry and Biochemistry, UCLA, Los Angeles, California 
90095-1569}

\date{\today}

\begin{abstract}
\asfsix undergoes two phase transitions upon cooling from 300~K. At \tco=103~K a 
charge-ordering (CO) occurs, and at \tsp($B$=9~T)=11~K the material undergoes a spin-Peierls 
(SP) transition. Within the intermediate, CO phase, the charge disproportionation ratio is 
found to be at least 3:1 from \carbon\ NMR \invtone\ measurements on spin-labeled samples. 
Above \tsp\, up to about 3\tsp\, \invtone\ is independent of temperature, indicative of 
low-dimensional magnetic correlations. With the application of about 0.15~GPa pressure, \tsp\ 
increases substantially, while \tco\ is rapidly suppressed, demonstrating that the two 
orders are competing. The experiments are compared to results obtained from calculations on the
 1D extended Peierls-Hubbard model.


\end{abstract}

\maketitle

Inhomogenous charge and spin structures are a consequence of
 competing interactions and therefore of general interest in correlated electron systems. 
 Examples include the high-$T_c$ cuprates \cite{Tranquada1995} 
and manganites \cite{Mori1998} as well as the quasi-2D organic conductors
\cite{Miyagawa2000}.
The quasi-1D salts made from TMTTF or TMTSF molecules are also susceptible 
to charge-ordered states. Independent of that, they are well-known for the sequence of 
ground states accessible 
by applying pressure or selecting different counterions. For example, the material \spf\ 
undergoes transitions from spin-Peierls, antiferromagnetic (AF), spin-density wave (SDW), 
and finally to superconducting (SC) ground states as the pressure is increased to 4-5~GPa 
\cite{Kobayashi2000,Wilhelm2001}. For a long time, it was known that another phase transition 
occurs in a number of TMTTF salts with both centrosymmetric ({\it e.g.}, AsF$_6$, SbF$_6$) and 
non-centrosymmetric ({\it e.g.}, ReO$_4$) counterions. Only recently \cite{Nad2000,Chow2000} 
was the broken symmetry associated with this transition identified as a charge 
disproportionation.

In TMTTF salts the characteristic temperature of the onset of the charge-ordered (CO) 
phase is high, on the order of 100~K. It indicates that the interactions driving 
the CO are relatively strong, and therefore potentially impact the electronic and magnetic 
properties of the disordered phase. Issues associated with CO correlations in these systems 
take particular 
relevance when considering that the nature of the metallic phase of TMTSF salts remains 
controversial \cite{Jerome1995b,Vescoli1998,GMihaly2000}. Below we report the results of 
a number of NMR measurements on \carbon\ spin-labeled samples of \sasf\ in the CO phase.  
Our principle result is a mapping of the temperature/pressure phase diagram of the SP and CO phases that 
includes a tetracritical point with a region of coexistence of the two forms of order. 
There is good agreement between the experiments and the results of calculations on the 1D extended 
Hubbard \cite{Seo1997} and Peierls-Hubbard models \cite{Mazumdar2000,Clay2001}.

A review of the characteristics of the CO phase and the phase transition is in order. With 
counterions PF$_6$, AsF$_6$, and SbF$_6$, the ordering temperature is 62~K, 103~K, and 154~K, 
respectively. Upon cooling, the salts made with the first two are already well into a region 
of thermally activated resistivities when the CO transition occurs. The last one undergoes a 
continuous metal/semiconductor transition at \tco. The mystery of the order parameter arose 
from the fact that no evidence for a superlattice was ever found with X-ray scattering. 
Charge order was 
identified as the proper description of the order parameter from NMR studies \cite{Chow2000}. 
At high temperatures, the unit cell consists of two equivalent TMTTF molecules related by 
inversion about the counterion \cite{Pouget1996}. A low-frequency divergence of the real 
part of the dielectric
susceptibility $\chi_e$ is consistent with mean-field-like ferroelectric behavior 
\cite{Nad2000,Monceau2001}. 
The observations were taken as evidence for a breaking of the inversion symmetry within 
the unit cell, and the spontaneous dipole moment is associated with the charge imbalance 
on the two molecules.

All of the experiments were performed in $B_0=9$~T magnetic field (\carbon\ NMR frequency  
$\nu$=96.4~MHz). Pressure was applied 
using a self-clamping BeCu cell, with Flourinert 75 serving as the pressure 
medium. The reproducible low-temperature pressure was calibrated in separate runs by measuring 
inductively the change of the superconducting transition temperature of lead. The NMR coil was 
constructed so that the molecular stacking axis (\a) was oriented normal to the static magnetic 
field. Any significant 
reorientation about \a\ between experimental runs is excluded by noting that the \carbon\ 
internuclear dipolar coupling remained unchanged.

\begin{figure}[htb]

\includegraphics[width=3.5in]{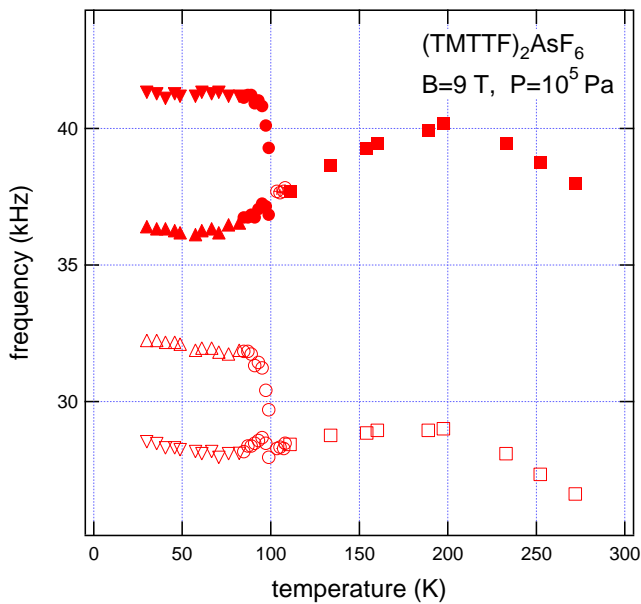}

\caption{Positions of the \carbon\ NMR peaks versus temperature at the magic angle at ambient 
pressure for \asfsix. The CO transition takes place at \tco=103~K.}

\label{fig:PeaksvsT}

\end{figure}

In Fig.~\ref{fig:PeaksvsT} we show the temperature dependence of the relative hyperfine shifts 
for all of the unique \carbon\ sites. Once the molecules are inside the solid, the two sites 
forming the bridge are inequivalent and have different hyperfine shifts. At high temperatures 
the two molecules within the unit 
cell are equivalent. Below the phase transition at \tco=103~K, each of the two lines split into 
two peaks with equal absorption strengths. We showed this to be a result of a charge 
disproportionation developing between two inequivalent molecules, and the order parameter 
amplitude is proportional to the difference of the NMR frequencies within each of the split 
lines. The character of the transition is 2nd order.

\begin{figure}[htb]
\includegraphics[width=3.5in]{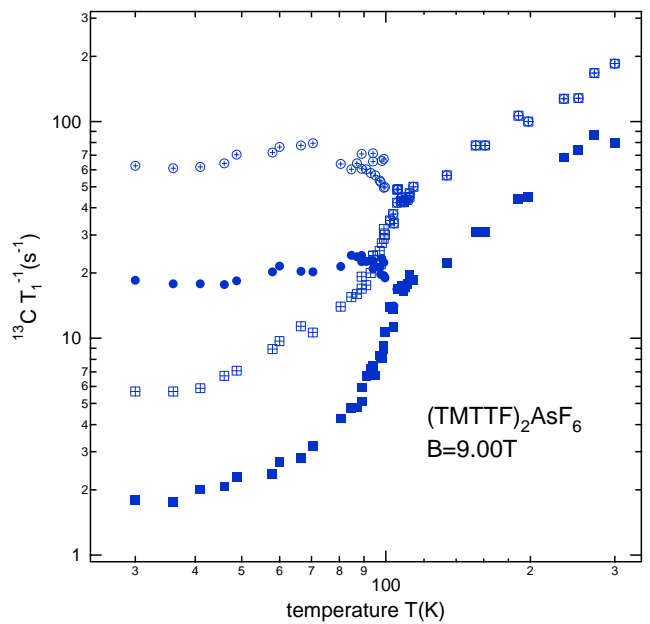}
\caption{$^{13}$\invtone\ relaxation rate as a function of temperature 
at ambient pressure in \asfsix. Spectra were taken at the magic angle,
the magnetization recovery was obtained by integrating over the 
distinct peaks of the absorption spectra. Similar markers represent
$^{13}$\invtone\ of peaks evolving from the same high temperature peak.}
\label{fig:T1vsTambient}
\end{figure}

The amplitude of the charge disproportionation can be estimated from the spin lattice 
relaxation rates. These are shown in Fig.~\ref{fig:T1vsTambient}. The 
relaxation rates are about one order of magnitude faster on one of the two types of molecules. 
Ideally, we expect that $T_1^{-1}$, dominated by hyperfine coupling, is proportional to the 
electronic density on a particular molecule. For isotropic hyperfine interaction and no 
spectral overlap whatsoever between the respective absorption lines, \iTone$\propto\rho^2$, 
with $\rho$ the molecular charge count. Neither one of these conditions is strictly true here, so 
the ratio of the relaxation rates of sites on the high- and low-density molecules determines a 
{\it lower bound} for the disproportionation, namely 3:1.

\begin{figure}[htb]
\includegraphics[width=3.5in]{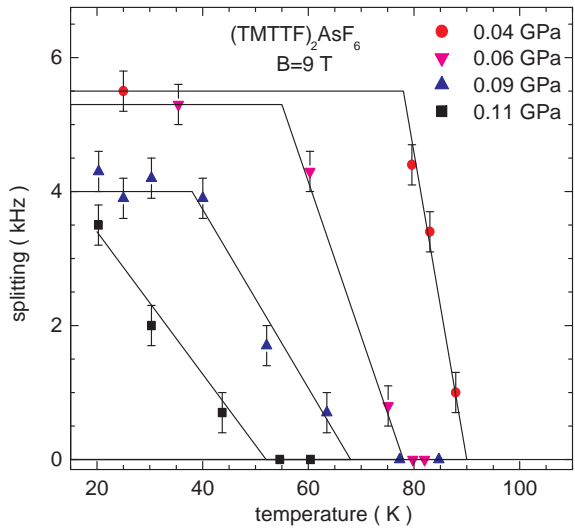}
\caption{The order parameter as a function of temperature at various pressures. 
The solid lines are guide to the eye.}
\label{fig:OPvsp}
\end{figure}

Modest applied pressures strongly depress \tco. The order parameter of the CO phase, taken 
from the temperature dependence 
of the splitting of the NMR lines, is shown in Fig.~\ref{fig:OPvsp} at different pressures. 
Three effects are visible as the pressure is increased: 1) The maximum splitting is decreased. 
2) \tco\ is decreased. 3) The order parameter develops much less steeply.

\begin{figure}[htb]

\includegraphics[width=3.5in]{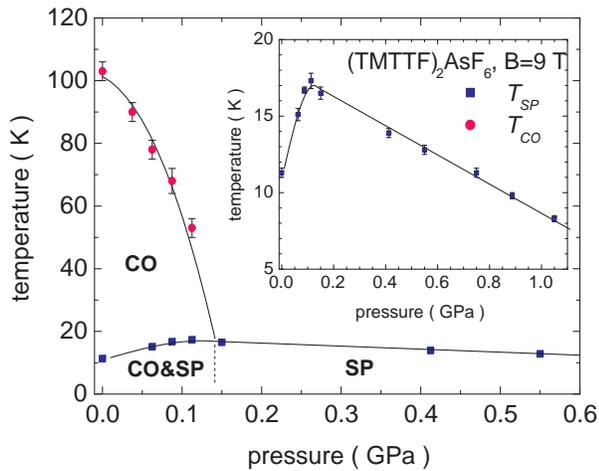}

\caption{Phase diagram of \asfsix established from NMR experiments at $B$=9~T. The lines are a
guide to the eye. The dashed line is used only to emphasize that a region of coexistence 
is present.}
\label{fig:phasediagram}
\end{figure}

\tsp\ is identified from the temperature dependence of either \carbon\ or \as\ 
\iTone. \carbon\ is a spin $I$=1/2 nucleus for which \iTone\ falls precipitously below \tsp. \as, 
on the other hand, has spin $I$=3/2, so it couples to electric field gradient (EFG) fluctuations 
produced, for example, by lattice vibrations. The peak in $^{75}$\iTone\ at \tsp\ is naturally associated 
with the critical slowing down of the soft 2$k_F$ phonons \cite{BrownUnpublished}. 
The completed phase diagram for \asfsix appears in Fig.~\ref{fig:phasediagram}. 
The CO phase diminishes quite rapidly: $P=P_c\lesssim$0.15~GPa is enough to suppress it.

\begin{figure}[htb]

\includegraphics[width=3.2in]{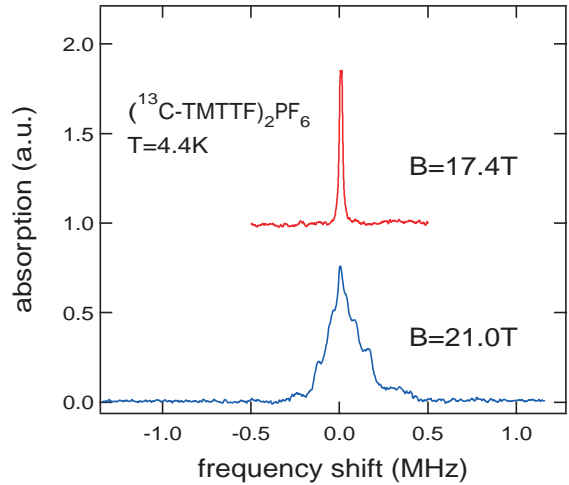}

\caption{\carbon\ NMR lineshapes of \spf\ at fields close to the critical field $B_c$,
at ambient pressure. The broadening for $B>B_c$ is characteristic of the incommensurate phase.}

\label{fig:ledges}

\end{figure}

The phase diagram has a remarkable feature: while \tco\ is decreasing, \tsp\ increases 
significantly up to about 150\% of its ambient pressure value. After the CO is suppressed at $P_c$,
 \tsp\  decreases weakly with additional pressure. The 
maximum of \tsp\ at $P_c$ indicates that the CO site order and the SP bond order are 
competing. If the competition is sufficiently weak,
the order parameters will coexist. Otherwise, a first-order transition line divides the two 
phases. In principle, the CO and SP order parameters could be simultaneously characterized 
by X-ray scattering experiments. To date, we are not aware of a published X-ray report of the 
CO phase, even at ambient pressure. We call attention to this not only because NMR is a 
local probe, but also because the methods desbribed above for determining the CO order 
parameter do not work in the SP phase. This is because the paramagnetic shifts are nearly 
absent for the singlet ground state and all \carbon\ sites are equivalent. 
There is an exception: magnetic fields $B$ larger than a critical value $B_c$ 
produce an incommensurate (I) phase, through the generation of triplet excitations. In a 1D 
picture, the triplets consist of soliton/anti-soliton pairs. Typically $B_c\propto T_{SP}$. 
Although it is not measured for \sasf, we know that $B_c$=19~T for the PF$_6$ salt 
\cite{Brown1998b}. Applied fields greater than 19~T resulted in broadened NMR lines associated 
with the staggered spin density of the excitations \cite{Brown1999b}. 
Lineshapes demonstrating the effect near to B$_c$ are shown in Fig.~\ref{fig:ledges}. 
At low fields, we see a single line showing that all \carbon\ sites are equivalent in the 
SP phase. If the CO is still present, we expect the 
absorption at high fields to consist of exactly 4 contributions all of which are identical. Two 
contributions are expected without the CO. Four {\it well-separated} "ledges" are evident on each 
side of the maximum at the higher field, which we interpret as coming from the nuclei near 
the center of the soliton-like excitations where the staggered spin density is maximum 
\cite{Hijmans1985,FagotRevurat1996}. The substantial differences in hyperfine fields on the four 
sites indicates that the disproportionation remains large even in the ground state, {\it i.e.}, 
the two orders coexist as shown in Fig.~\ref{fig:phasediagram}.
A smaller charge disproportionation is also expected in the SP phase \cite{Clay2001} at high pressures,
but our experimental setup did not allow to investigate it.

We should emphasize that the 1D extended Peierls-Hubbard model 
contains most of the essential physics that produces the observed phase diagram. 
Justification for a quasi-1D point of view follows from the fact that the molecular spacing 
along the stack alternates \cite{Pouget1996}. It can be modeled  
by introducing intrachain hopping integrals $t_1$ and $t_2$ along the stack. 
Therefore, the system is half-filled and electron-electron Umklapp scattering in 
1D opens a charge gap $\Delta_{\rho}$ \cite{Emery1982}. Provided that 
$\Delta_{\rho}\gg t_{\perp}$, where $t_{\perp}$ is the transverse hopping, 
1D confinement leads to an insulating state  even at high temperatures and at $T>T_{CO}$.  

Without phonon coupling, the 1D extended Hubbard model will produce a CO state if the Coulomb 
interaction is sufficiently large and long-ranged \cite{Emery1982}, {\it i.e.}, the nearest-neighbor 
interaction $V>V_c$ \cite{Seo1997}. In the TMTTF salts, it is expected that $V$ is very 
close to the predicted threshold obtained in strong-coupling perturbation theory 
\cite{Mila1995,Clay2001}. Further, the large lower bound that we infer for the charge 
disproportionation is quantitatively reasonable when compared to model parameters \cite{Seo1997,Clay2001}. 
The principle effect of pressure is to increase the hopping integrals, which  
leads to a rapid destabilization of the CO phase, just as we observe. $P>P_c$ results in a 
SP ground state, once the phonon coupling is included. 

Regarding the high-temperature CO phase, there are two qualitative differences between  
recent numerical works and our experimental observations. One of them is the
expectation that the CO occurs together with the metal-insulator transition \cite{Clay2001}.
The other is the character of the transition.
The measurements are consistent with the CO phase transition being a continuous 
one along the entire line, whereas {\it mean-field} calculations describe the 
transition as first order. Including the dimerization of the 1D stack is known 
to eliminate the first-order character in the mean-field calculations of the 
1D extended Hubbard model \cite{Seo1997}. 
Therefore, it is very likely that the dimerization is crucial in {\it three} respects. 
First, it confines coherent charge motion to the stacks and because of this it is appropriate
to compare to 1D models. Second, the confinement assures that a metal-insulator crossover 
occurs at a high temperature, perhaps far above \tco. Third, it changes the character of the 
CO transition from first-order to second-order. 

Returning to the issue of metal vs. insulator in the TMTTF salts, there is one choice of 
centrosymmetric counterion, SbF$_6$, which undergoes a real transition to an insulating
state at \tco\ \cite{Laversanne1984}. The reason is thought to be that the larger size 
of SbF$_6$ relative to, say PF$_6$, leads to a smaller difference between the alternating
hopping integrals, $t_1$ and $t_2$, and the 1D Umklapp scattering is not strong enough
relative to $t_{\perp}$ to open the charge gap. It still undergoes a CO transition at 
\tco=154~K, though the system should be thought of as inherently quasi-2D. 
Consequently, we do not expect the ground state to have spin-Peierls character, 
and indeed, it is antiferromagnetic \cite{Coulon1986}.

In conclusion, we studied the pressure dependence of the relative stability of the CO and SP 
phases observed in \asfsix\ using \carbon\ NMR spectroscopy and spin-lattice relaxation 
measurements. The coexistence of the two orders was established along with the existence 
of a tetracritical point in the temperature/pressure phase diagram. The CO phase is suppressed
easily with modest pressure; the natural interpretion is that it results from the effect of the 
increasing bandwidth relative to the strength of the near neighbor Coulomb repulsion $V$ within 
the stacks. The lower-bound for the charge disproportionation is set at 3:1. 
The phase diagram and disproportionation amplitude are consistently compared to results 
of calculations on the 1D extended Hubbard and Peierls-Hubbard models. 

\small ACKNOWLEDGEMENTS. The work was supported by the NSF under grant no.~DMR-9971530.
We would like to acknowledge helpful discussions with S.~Brazovskii, R.~T.~Clay, 
S.~Kivelson, and S.~Mazumdar.

\end{document}